\documentclass[namedreferences]{kluwer}
\usepackage[dvips]{epsfig}
\def\nd{\noindent}
\begin{document}
\begin{article}
\begin{opening}

\title{MULTI-WAVELENGTH OBSERVATIONS OF AN UNUSUAL IMPULSIVE FLARE ASSOCIATED WITH CME}
\author{WAHAB \surname{UDDIN}$^1$, RAJMAL JAIN$^2$, KEIJI YOSHIMURA$^3$, RAMESH CHANDRA$^1$, 
T. SAKAO$^3$, T. KOSUGI$^3$, ANITA JOSHI$^1$ and M. R. DESPANDE$^2$}

\institute{$^1$Aryabhatta Research Institute of Observational Sciences,  Manora Peak, 
Nainital-263 129, India (Formerly State Observatory) (e-mail: wahab@upso.ernet.in)}
\institute{$^2$Physical Research Laboratory, Ahmedabad - 380 009, India}
\institute{$^3$The Institute of Space And Astronautical Science, Sagamihara 229 Japan}

\begin{ao}
Dr. Wahab Uddin\\
Aryabhatt Research Institute of Observational Sciences \\
(Formerly State Observatory)\\
Manora Peak, Naini Tal$-$263 129,\\
INDIA.\\
Phone: +91-05942-235583, 235136\\
Fax: +91-05942-235136\\
email: wahab@upso.ernet.in
\end{ao}

\begin{abstract}
We present the results of a detailed analysis of multi-wavelength observations of a 
very impulsive solar flare 1B/M6.7, which occurred on 10 March,
2001 in NOAA AR 9368 (N27 W42). The observations show that the flare is very impulsive with very hard spectrum in HXR that reveal non-thermal emission was most dominant. On the other hand this flare also produced type II radio burst and coronal mass ejections (CME), which are not general characteristics for impulsive flares. 
In H$\alpha$ we observed the bright mass ejecta (BME) followed by drak mass ejecta (DME). Based on the consistence of the onset
times and direction of BME and CME, we conclude that these two phenomena are closely associated.
It is inferred that the energy build-up took place due to 
photospheric reconnection between
emerging positive parasitic polarity and predominant
negative polarity, which resulted as a consequence of flux cancellation. 
The shear increased to $>$80$^o$
due to further emergence of positive parasitic polarity causing strongly enhanced cancellation of
flux. It appears that such enhanced magnetic flux cancellation in a strongly sheared region triggered 
the impulsive flare.
   
\end{abstract}
\end{opening}

\vspace*{-0.8cm}
\section{INTRODUCTION}
 Impulsive flares produce a wide range of emission from $\gamma$-rays, X-ray, EUV, visible,
microwave (MW) and longwave radio emission (Miller et al., 1997). The X-ray emission is comprised
of thermal, superhot and non-thermal components. Based on the classical nonthermal thick-target
model it is proposed that the electrons accelerated to $>$ 100 keV in the corona spiral downwards,
and creating MW (Brown, 1971; Lin and Hudson, 1976; Kundu and White, 2001). When they
reach the footpoints of a coronal loop, they produce hard X-ray (HXR) and drive evaporation, which
fills the loop with hot plasma emitting in soft X-rays (SXR).
      
    For better understanding of the various aspect of flare phenomena at different height of
solar atmosphere, it is important to resolve the  different structures involved and temporal
evolution during the flare. Therefore simultaneous  observations of chromospheric (optical) coronal
(X-ray and MW) with high temporal and spatial resolutions can provide wealth of information
for flare diagnostic.

 In this paper we present analysis of 10 March 2001 impulsive flare at different wavelengths, viz.,
H$\alpha$, HXR, SXR and radio wave to understand the various physical processes taking place at different
height in the above flare. 

\vspace*{-0.6cm}
\section{OBSERVATIONS}

In the peak of solar cycle 23 on March 10, 2001 we observed a very impulsive solar flare 1B/M6.7
in H$\alpha$ from NOAA 9368 at the location N27 W42.
The flare observations were carried out at State Observatory, Manora Peak, Nainital, India with 
15 cm, f/15 Coud\'{e} Solar
Tower Telescope equipped
with  Bernhard Halle H$\alpha$ filter and Wright Instrument CCD camera system ( 16 bit, 385$\times$576
pixel, pixel size = 22 micron square). With the help of barlow lens the image has been magnified
twice, so we get resolution of 1$^{''}$ per pixel.
The observed filtergrams were corrected
using dark current and flat field images taken through CCD during the observations. All images were
re-registered.
The flare has been simultaneously observed by Yohkoh/HXT (Kosugi et al. 1991) and  SXT 
(Tsuneta et al. 1991). To study the flare in MW we have used
Nobeyama, Japan observations.
SOHO/LASCO (Brueckner et al. 1995) and SOHO/EIT (Delaboudiniere et al. 1995)
observed the CME associated with the present flare under study. 
This flare also produced several strong radio bursts
and it was well observed by the HiRAS observatory, Japan and WAVES/WIND (Bougeret et al. 1995)
spectrometers in a broad range of radio frequencies.
To understand the energy build-up process we used SOHO/MDI (Scherrer et al, 1995) and NAOJ, Mitaka, Japan
magnetograms.

\vspace*{-0.6cm}
\section{ANALYSIS AND RESULTS}
\subsection{Temporal Evolution}

On 10 March 2001 a very impulsive flare occurred in the following sunspot region of
NOAA AR 9368 located at N27 W42 in heliocentric coordinates. GOES SXR observations marked it as a 
M6.7 class while in H$\alpha$ it was recorded as 1B class.
Shown in Figure 1 are light curves of the HXR emission from the flare as observed by HXT. 
The light curves are in the four energy windows: L (14$-$23 keV),
M1 (23$-$33 keV), M2 (33$-$53 keV), and H (53$-$93 keV). In all the four energy
channels the flare show that impulsive component is very strong having maximum

\begin{figure}[t]
\vspace*{-4.0cm}
\epsfig{file=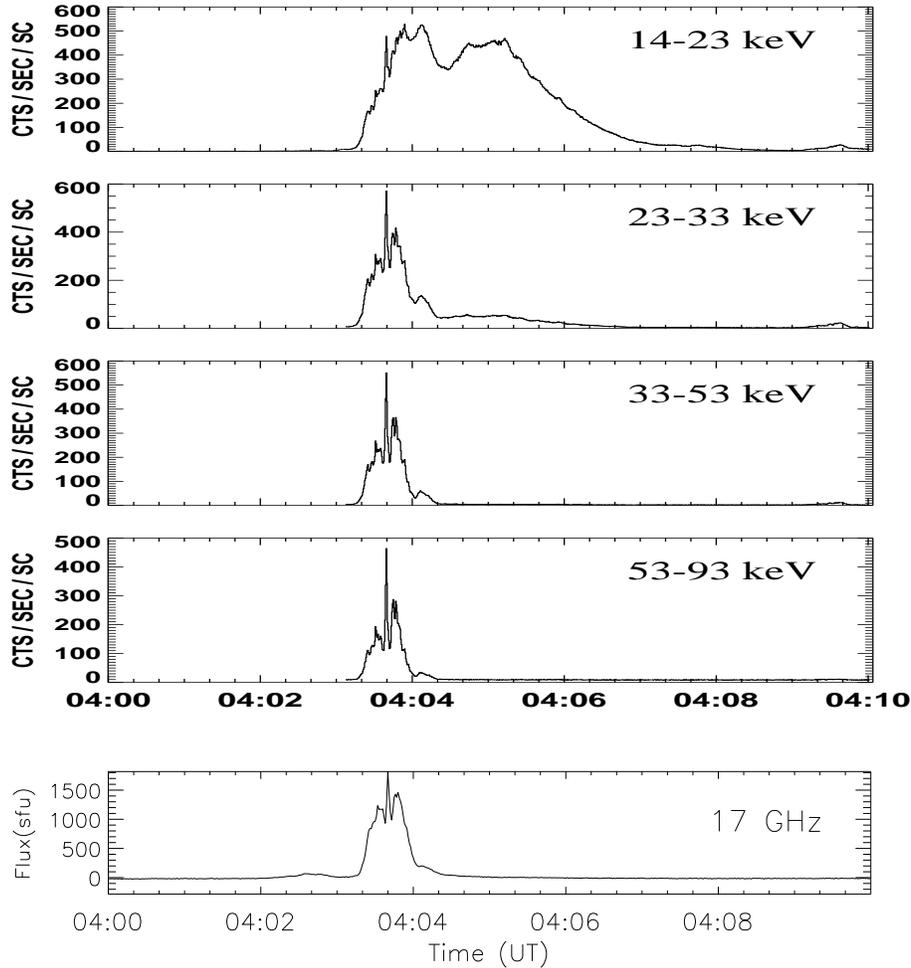, width=7.7in, height=11.0in}
\vspace*{-12cm}
\caption{Hard X-ray burst time profile of 10 March, 2001 flare taken with Yohkoh/HXT
in four energy bands L, M1, M2 and H. 
The bottom panel is microwave time profile 
at 17 GHz observed by Nobeyama radio observatory, Japan.}
\end{figure} 
            
\nd counts 528, 570, 550 and 463 in L, M1, M2 and H bands respectively. 
The HXT data show that the flare had a quite hard spectrum.
In order to summerizing the HXR observations according to the time profiles 
, we divide the flare into three phases: (a) precursor phase (04:03:00 - 04:03:15 UT),
(b) impulsive phase (04:03:15 - 04:03:50 UT) and (c) gradual phase ( after 04:03:50 UT).
In this event precursor phase is very short about 15 sec. However, the impulsive phase has
three separate spikes, which are clearly seen in all four energy bands. The impulsive
component refers to non-thermal emission.  The superhot and the thermal components
are visible after 04:03:50 UT, however only in L and M1 energy bands.
  
  The  bottom panel of Figure 1 presents evolution of the flare at 17 GHz.
observed with NoRP Nobeyama. The MW time profile shows very impulsive nature
similar to HXR at higher energy bands viz., M1, M2 and H bands respectively.
   
We derived emission measure and temperature from M1/L ratio of HXT data that is
10$^{49}$ cm$^{-3}$ and 32 MK respectively. In Figure 2 (top) we show the emission
measure and temperature variation as a function of time, while in Figure 2 (bottom)
photon index as it varies as a function of time. The emission measure as well as 
photon index indicate that the flare
component before 04:04 UT is purely nonthermal and after 04:05 UT it
was thermal. It falls exponentially till 04:08 UT by almost three orders.
This indicates that thermal plasma was radiating its heat very fast in the flare.
It appears that during the impulsive phase of the current flare high energy electrons
were  accelerated in a very short time.

The H$\alpha$ observations of the flare show two main kernels K1 and K2 (cf., Figure 7)
appeared around 04:02 UT, which release most of the flare energy. The H$\alpha$ light
curves of K1 and K2 are shown in Figure 3 (left), while in soft X-ray emission in Figure 3 (right).
This Figure reveals a gross correlationship between the chromospheric emission
( low temp  10${^4}$ K) and the coronal emission (high temp 10${^6}$ K).
The time profiles of K1 and K2 show sharp rise similar to HXR emission indicating
impulsive nature of the flare. The intensity of kernel K1 and K2 reached to maximum
at 04:04:45 and 04:04:05 UT respectively.
It may be noted from Figure 3 (right) that preflare thermal heating as seen by Yohkoh/SXR emission 
started around 04:00:51 UT, which suddenly and impulsively enhanced
around 04:03:30 UT, almost in simultaneous to H$\alpha$ and about 15$-$20 sec. later to HXR impulsive
onset time. The SXR emission of kernel A reach to maximum at 04:06:00 UT with ma- 

\begin{figure}
\vspace*{-3.0cm}
\hspace*{-2cm}
\epsfig{file=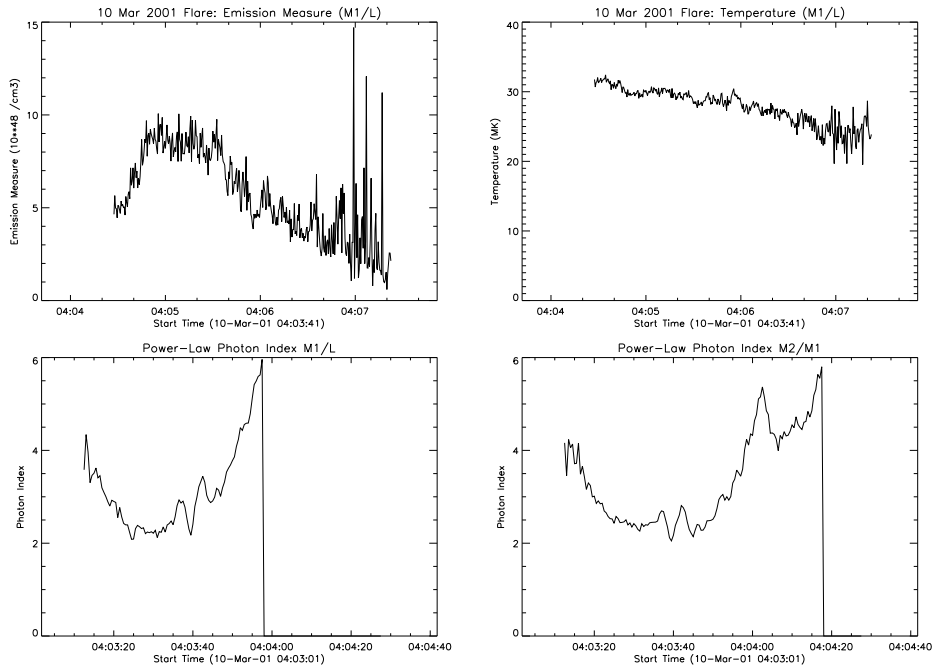,width=16.5cm,height=16.5cm}
\vspace*{-5.6cm}
\caption{ Temporal variation of emission measure and temperature (top) drived from M1/L 
ratio of HXT data and photon index variation (bottom) as a function of time.}
\end{figure}                                                                    
 
\begin{figure}
\vspace*{-0.40cm}
\epsfig{file=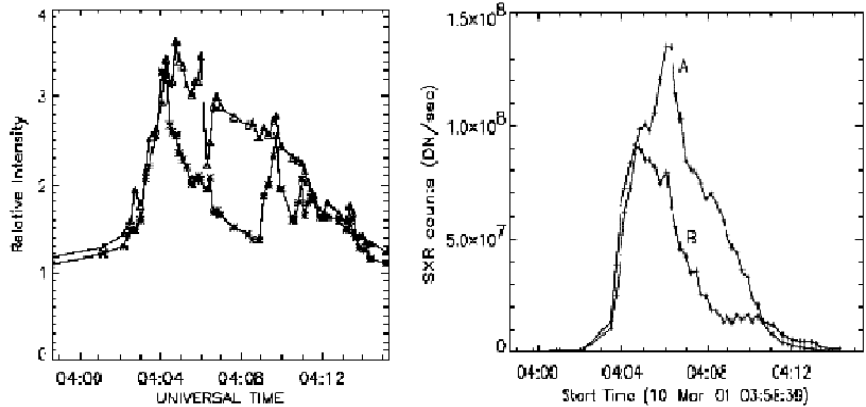,width=12.0cm,height=6.0cm}
\vspace*{-0.9cm}
\caption{Intensity time profiles of flare kernels K1 (triangle)(A) and K2 (asterisks)(B) 
in H$\alpha$ and Soft X-ray (Yohkoh SXT/AlMg)
respectively.}
\end{figure}                                                                    

\nd ximum count (DN/sec)
1.35$\times$10${^8}$ and kernel B reached maximum at 04:04:25 UT with maximum count 5.8$\times$10${^7}$.
In kernel B there are small fluctuations during the maximum phase. From time profiles of
H$\alpha$ and SXR we also found that the preflare heating started in H$\alpha$ at about
04:01:11 UT and in SXR at about 04:00:51 UT as measured with observing cadence limitations.
This 
indicates that in prior to impulsive phase chromosphere and corona are already intermittently
connected. However, the maximum intensity reached in H$\alpha$ and SXR at 04:04:45 and 04:06:00 UT
respectively. The impulsive phase transports the accelerated electrons very impulsively,
which in turn deposit their energy by collision with ambient material in lower corona and
chromosphere as seen in the present case as K1 and K2 in H$\alpha$ and corresponding A and
B bright SXR sources.
However, on the other hand, it is clear from the light curves of H$\alpha$,
SXR, HXR that the current flare was of very short duration, almost of 12 min.

\subsection {Dynamic Radio Spectra} 

     Type II burst always follow flares, and hence the associated shock must be flare related;
there remains a controversy as to weather they are driven primarily by CME or occur with the flare.
During this flare several type II (slow drift), type III (fast drift) and type V radio bursts have been observed.

The dynamic radio spectra observed from Hiraso spectrograph shows group of metric type III burst with maximum
intensity 3 at 04:01 UT. The metric type II burst began at 04:04 UT with starting frequency 400 MHz,
which is equivalent to heliocentric distance 1.05 R$\odot$.
The burst ended at 04:16 UT at 40 MHz frequency. We estimated the shock speed from this type II burst
spectrum of the order of 800 km sec$^{-1}$.

    On the other hand, WAVES/WIND observed interplanetary type II radio burst during  04:18$-$04:32 UT in
the frequency range 4000-14000 kHz, however with a weak fundamental and harmonic emissions.
A strong type III radio burst was also recorded around 04:03 UT by WIND.

\subsection{SOHO/LASCO and EIT Observations}

\noindent SOHO observed  a coronal mass ejection and the coronal dimming associated with the
current flare under study. Figure 4 illustrate a series of SOHO/LASCO C2 and C3 images of CME.
At first, the CME appeared in C2 as a bright patch at 04:26 UT, marked by arrow. In C3 it
appeared at a radial distance 9.19 R$\odot$ at 05:18 UT. In Figure 8 (right) we plot the
height of the leading edge of CME as a function of time. The average velocity of the CME
derived from this height-time plot is of the order of 800 km sec$^{-1}$, almost same to
the shock speed derived from type II radio burst.  The CME was moving ahead with
deceleration of about 23.2 m sec$^{-2}$. The angular width and position angle of the
CME were 81$^o$ and 310$^o$ respectively.

\begin{figure}
\vspace*{-0.4cm}
\hspace*{-0.3cm}
\epsfig{file=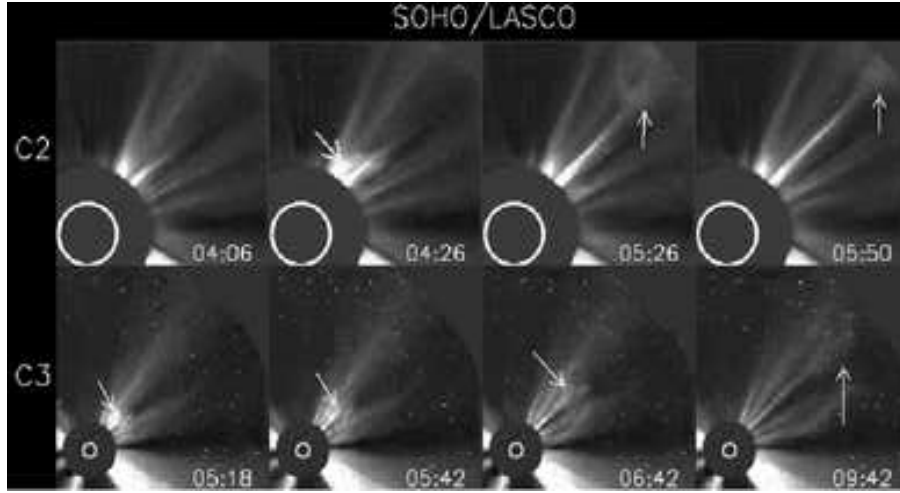,width=6.5cm,height=12cm,angle=-90}
\vspace*{-0.1cm}

\caption{Images from C2 and C3 LASCO coronagraph on SOHO. The observation times are 
indicated on each image. North 
in on the top and east to the left}
\end{figure}                                                                    
 
   SOHO/EIT at 195 \AA\ observed coronal dimming at the flare location associated with
the CME. The dimming occurred in
the north of the flare after 04:02 UT, i.e., after the onset of the flare. The evolution
timescale of coronal dimming is faster than typical radiative cooling timescales in
the corona, indicating that density depletion via expansion or ejection is most likely
responsible for the dimming process (Hudsonet al., 1996), which is in
agreement to present CME event.

\subsection{Morphology of the active region}

NOAA AR 9368 emerged on 2 March 2001 at N27 E48 location as a small spot of $\alpha$
type magnetic class. The active region slowly developed as $\alpha$$\beta$ type along with
an emerging magnetic flux of opposite polarity near main leading and following spots on 6 March.
On 7 and 8 March 2001 the region expanded as noted from Figure 5. In Figure 5, we 
present a sequence of H$\alpha$ filtergrams taken in line center and off bands, as well as the magnetograms
taken by SOHO/MDI, from 7$-$10 March 2001. The H$\alpha$ filtergrams on 7$-$8 March shows bright plages
and dark arches in the central part of the region (cf., Figure 5).

\begin{figure}
\vspace*{-0.5cm}
\hspace*{-0.3cm}
\epsfig{file=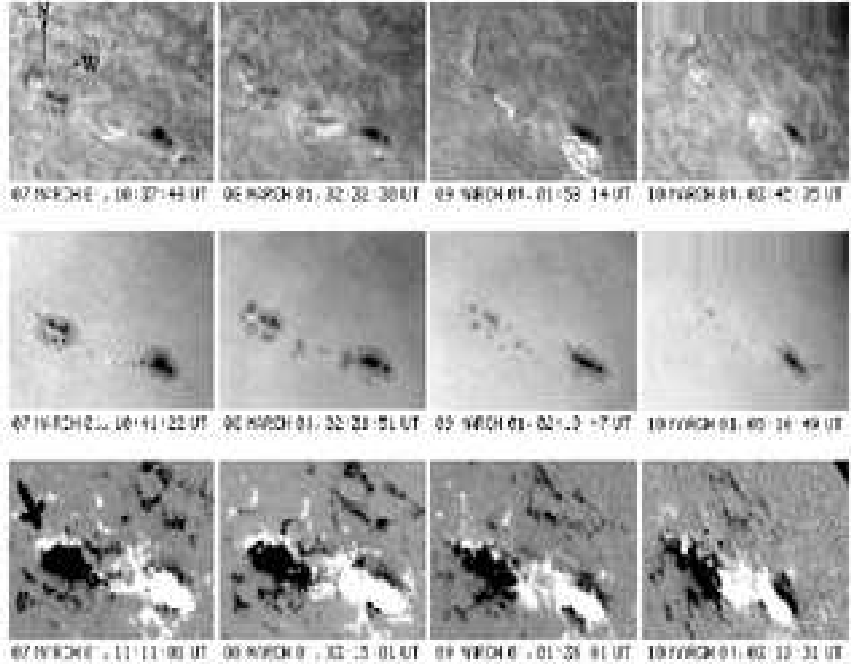,width=12cm,height=20cm}
\vspace*{-11cm}
\caption{Evolution of NOAA AR 9368 from March 7 to 10. Top and middle rows are H$\alpha$ 
center and off band 0.8 \AA\ taken at S.O. Nainital and the bottom row represents SOHO/MDI magnetograms.
The arrow in the figure represent the parasitic polarity. The field of view (FOV) of the images is 240$^{''}$$\times$195$^{''}$.}
\end{figure}                                                                    

The significant changes in active region occurred between 8 and 10 March 2001.
On 8th March the following spot began to fragment and continued on 9th and 10th March.
This activity was further accompanied by flux emergence of positive polarity.
On the other hand, the leading spots did not show any considerable changes. However,
on 9 March 2001 a 1N/M1.5 class impulsive flare occurred near the leading spots (cf., Figure 5).
It was noted by us that the active region had developed into $\beta$$\gamma$ class on 9 March, which 
continued on 10 March 2001.

The  SOHO/MDI magnetograms and intensitygrams from 2 to 10 March 2001 of the active region show
that the following spot was growing faster than leading spot, however it decayed also faster.
The magnetograms of the active region (cf., Figure 5) show that positive parasitic polarity
emerged near the following spot. This positive flux (MMF) moved towards negative polarity
dominated following spot resulting in flux cancellation. 
The flux cancellation rate between 8 and 10 March 2001 at the flare site was of the order
of 10${^{19}}$ Mx hr${^{-1}}$.
It seems that this MMF and the
process of flux cancellation by it was a major source for energy build-up through
photospheric reconnection. In fact at the location of such flux cancellation a filament
is formed in chromosphere suggesting the set up of reconnection process at low lying levels
(subphotospheric to photospheric level). 

    The high flux cancellation and thereby increasing shear in the following part of the
active region from 8 to 10 March 2001 showed development of dark arches and twisted
sigmoid type filament in H$\alpha$. This indicates the storage of magnetic energy at
the slow rate. A small and almost straight H$\alpha$ filament appeared on 9 March, which changed
into the sigmoid structure (length about 26000 km, width 3500 km) on 10 March before flare onset
(cf., Figure 7 ). 
 
   Vector magnetogram taken at NAOJ, Mitaka, Japan on 10 March 2001 before flare (cf., Figure 6) 
shows the magnetic field 
topology at the flare location. The flare occurred at the positive flux emergence site where the magnetic shear
was high $>$ 80$^o$ (cf., Figure 6).

\subsection{H$\alpha$ Morphology of the flare}

 The current impulsive flare (1B/M6.7) under study that occurred on 10 March 2001 began in
H$\alpha$ at around 04:01 UT. A total 130 H$\alpha$ 
filtergrams have been analysed for our
present study. In Figure 7 we present a sequence of a few selected H$\alpha$ filtergrams,
which show significant 

\begin{figure}
\vspace*{-0.4cm}
\epsfig{file=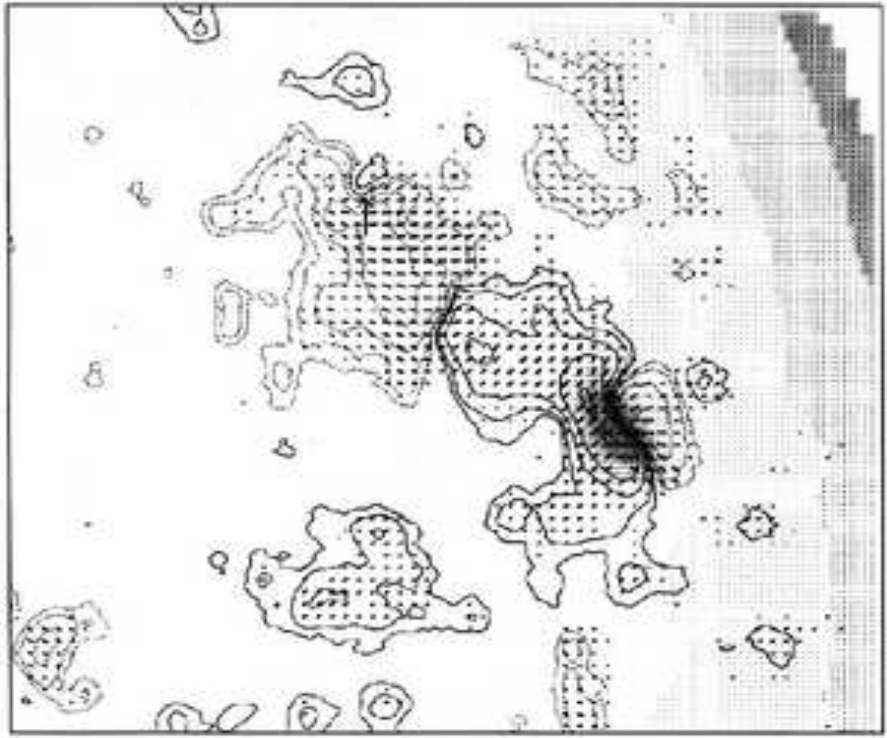,width=12cm,height=10cm}
\vspace*{-0.6cm}
\caption{The vector magnetogram (before flare) of the active region 
observed by NAOJ, Mitaka, Japan on March 10, 2001 at 00:10:18 UT. 
The contour levels are +/- 10, 20, 50, 100, 200, 500 and 1000 gauss (Solid =plus, dotted=minus).
High shear at the flare location is shown by arrow. North is top and east to the left. The FOV of the magnetogram is 340$^{''}$ $\times$320 $^{''}$.} 
\end{figure}

\begin{figure}
\vspace*{-0.4cm}
\hspace*{-0.3cm}
\epsfig{file=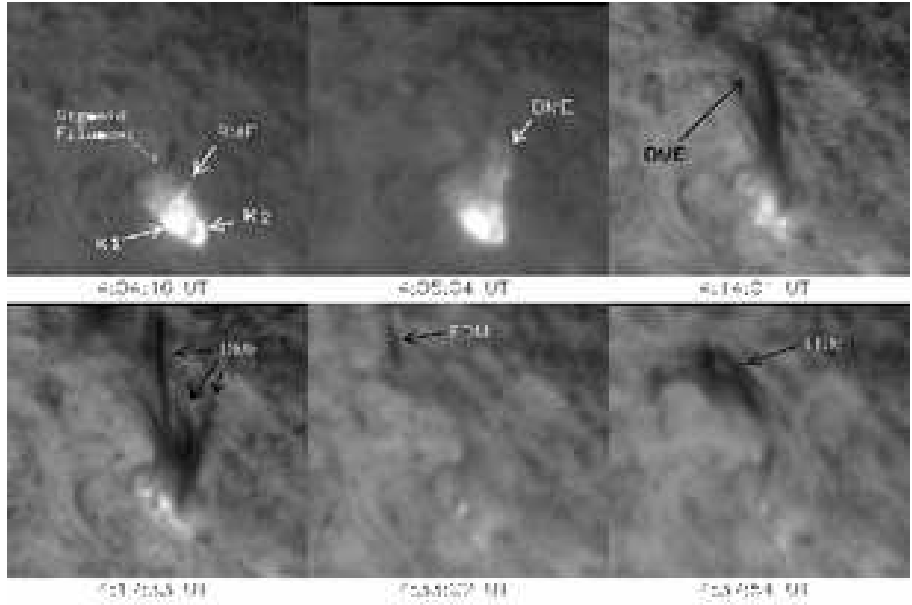,width=12cm,height=8cm}
\vspace*{-0.5cm}
\caption{Selected H$\alpha$ filtergrams of 1B/M6.7 class impulsive flare on March 10, 2001.
During maximum and decay phase of the flare the jets like BME
DME and the condensation of filament dark material (FDM) are shown by arrows.
North in on the top and east to the left. The field of view (FOV) of the images is 160$^{''}$$\times$145$^{''}$.}
\end{figure}                                                                    

\nd changes during flare evolution. As briefly described in section 3.1,
the H$\alpha$ flare began as a two small bright kernels, defined as K1 and K2, which rapidly
expanded in a bright oval shape during the flare peak phase. Our H$\alpha$ observations
showed remarkable bright and dark mass ejection activity during the flare. The BME started along 
with DME from kernel K1 and K2 around maximum phase at 04:04:16 UT, which 

\begin{figure}
\vspace*{-8.5cm}
\hspace*{-2.5cm}
\epsfig{file=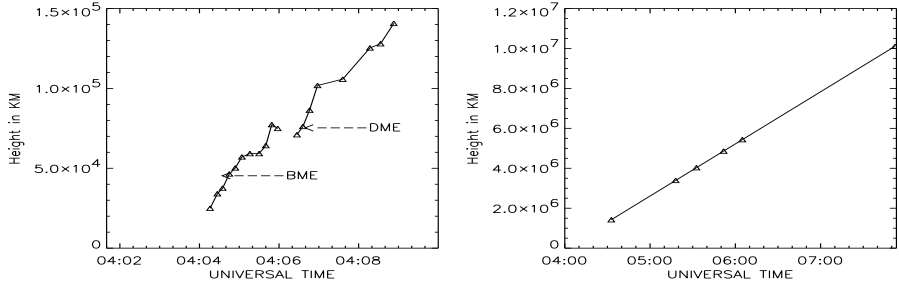,width=16cm,height=20cm}
\vspace*{-8.6cm}
\caption{Height Time plot of BME, DME (left) and CME (right)}
\end{figure}                                                                    

\nd continue until 04:05:05 UT (cf., Figure 7). The average sky plane
speed of the BME is estimated to 600 km sec$^{-1}$. However, later on around 04:06:58 UT
only DME was seen as a bunch of surges and jets
 from kernel K1 and K2 . The average sky plane speed of DME was about 400 km s$^{-1}$.
The DME also triggered mass ejection from nearby but opposite side of the sigmoid filament.
We noted formation of dark loop/arch filaments system in the active region onwards 04:24:50 UT.
We infer from the H$\alpha$ filtergrams that it  is due to connection between both sides of
dark mass ejection, perhaps as a consequence of condensation and hence downwards motion of
the material in the pre-existing loops. The loop like arch filament system is clearly
visible at 04:33:02 and 04:37:54 UT (cf. Figure 7). However, it faded out rapidly after 04:41:06 UT.

   On the other hand, the flare showed fast changes during its evolution. The bright kernels K1
and K2 were the main centers of energy release. The H$\alpha$ filtergrams showed that kernel
K1 consisted of large number of twisted flux tubes/loops in contrast to kernel K2.
Detailed analysis of our H$\alpha$ filtergrams/movie of BME and DME showed that the solar material
(plasma) ejected through the helically twisted flux ropes, which get detwisted
during mass ejection. This indicate the relaxation of sheared magnetic field.
From the above scenario it is unambiguously clear that the current  impulsive but compact
flare was associated with a blast wave of which the BME/DME were part and remnants.

\vspace*{-0.7cm}
\section{DISCUSSION}

The NOAA AR 9368 began as a very simple $\alpha$ magnetic class region, and became magnetically
complex in preview to emerging magnetic fluxes of opposite polarities that causing cancellation
of magnetic fields and also increased the shear as seen on 10 March 2001.  The 1B/M6.7 class
impulsive flare occurred in this highly complex evolving magnetic field with $\beta$$\gamma$ magnetic
class, and at the location where fields were highly sheared. 
From 8-10 March 2001
it considerably decayed and fragmented into many small spots. This kind of variation in
following spot group appears to be a result of significant flux cancellation that was going
on since 8 March due to appearance of MMF in the vicinity of following spot. We propose
that such enhanced flux cancellation was the basic physical process for energy build-up
and storage through photospheric reconnection, which, however, was inferred from the
thin filament formation. Our these observations are in agreement to Litvinenko, and Martin (1999).
However, on the other hand,  the leading spot still maintained its large size.
The energy build up through cancellation of magnetic flux by MMF is also earlier
suggested by many investigators (Uddin, et al., 1986, Kurokawa, 1987, Kosovichev and 
Zharkova, 2001).
   
     Mass motion play an important role in solar flares (Martin, 1989).
Figure 8 shows the height-time plots of the BME, DME (Left) and CME (right).
The BME observed around 04:04:16 UT from kernel K1, which found associated with CME  and
type II radio bursts in time and velocity scales. On the other hand our observations reveal
that the DME during their motion in the various flux tubes were
getting detwisted indicating relaxation of magnetic field  in the flaring region.
The estimated speed of the DME was about 400 km sec$^{-1}$, which was lower than the escape velocity
in the chromosphere and thereby they were appearing moving in the flux tubes controlled
under the magnetic force $B^2$/8$\pi$.

   The flare studied here is unusual in the sense though it is compact and impulsive in all wavelengths
but was associated
with a fast CME and Type II radio bursts, which otherwise found in association to large eruptive or
long duration event (LDE) type  flares. Such flares "confined but eruptive" (Moore, 1991) are of
specific importance as they provide unique opportunity to understand the physical processes that
operate for energy build-up and release impulsively following to a fast mass ejection. It appears
that in the present flare a large scale restructuring was going on though the flare appeared as
compact in size. We estimated the loop length between the EFR of opposite parasitic positive
polarity and south following spot of about 10$^4$ km. It appears that as soon as the reconnection
started the accelerated electrons reached to the foot points in a fraction of a second even if we
consider the speed between 0.5$-$0.8c. On the other hand, as mentioned earlier, the reconnection
rate was very high in view of fast speed of CME (Shibata, 1995, 1998, 2001), which made the
flare most impulsive, and hence the electrons were moved fast in the loops directing towards
the foot points where they produced bright kernels K1 and K2 through thermalization of their energy.
The fast moving electrons produced non-thermal HXR emission during their passage in the loops by
collision with the ambient material inside the lower corona and chromosphere. The strong MW emission
observed in simultaneous with HXR and H$\alpha$ strongly suggest that the same population of
electrons might have produced them through gyrosynchrotron process. We estimated electron density
above 10$^{{13}}$ cm$^{-3}$ that produced impulsive HXR, MW and H$\alpha$ kernel K1. We
interpret strong MW emission (Chandra et al., 2004) seen at K1 location because of significantly
high magnetic field of the order of 500 gauss at EFR of north polarity at parasitic site as
recently reported by Asai et al. (2002).
This flare is a electron rich event showing narrow ejecta (St.Cyr et al., 2001) and classified as 
white light flare of class I (Liu et al., 2001)

   The current impulsive flare event associated with CME appears to be explained better by
unified model (Plasmoid-induced reconnection model) of solar flares by Shibata et al.,
(1995, 1998, 2001), breakout model of Antiochos (1998), Sterling (2001) and also 
by Tether-cutting model (Sturrock, 1989; Moore et al, 2001; Sterling and Moore, 2003).

\vspace*{-0.7cm}
\section{CONCLUSION}

NOAA AR 9368 appeared on 02 March 2001 as most simple $\alpha$ type magnetic class region,
which slowly became magnetically complex as a function of time. Several MMF of positive polarity
were observed near following spots group  of south dominating polarity and causing cancellation
of magnetic field. It appears that magnetic field cancellation at following spot side as well
as increasing shear was the basic physical process for energy build-up.
The high shear, and hence the enhanced magnetic stress erupted the flare,
which was associated with the CME through a shock produced by it. The fast passage of CME induced the
reconnection, which in turn was responsible for the impulsive energy release as observed in the
flare and associated HXR and MW emission.
It seems that compact and confined impulsive flare
associated with CME require continual appearance of opposite polarity emerging fluxes (MMF)
and increasing shear for impulsive energy release. 

\begin{acknowledgements}

\noindent This work has been carried out under India Japan Cooperative Science Program (IJCSP),
which is supported by Department of Science and Technology (DST), Govt. of India and Japan
Society for Promotion of Science (JSPS). 
We express our sincere thanks to Prof. Satoshi Masuda,
Prof. B. V. Somov and Prof. T. Sakurai for very useful discussions, which helped in the 
interpretations of our results. We are also thankful to the referee for constructive comments and suggestions. 

\end{acknowledgements}

\end{article}
\end{document}